\documentclass{article}
\usepackage{amsmath}
\usepackage{amssymb}
\usepackage{graphicx} 
\usepackage{setspace} 
\usepackage{microtype} 
\usepackage{booktabs}
\usepackage{caption}
\usepackage{tabularx}
\usepackage{arydshln}
\usepackage{float} 
\usepackage{multirow}
\usepackage{xcolor}
\renewcommand{\arraystretch}{1.0} 

\usepackage{geometry} 
\begin{document}

\title{Volatility Forecasting in Global Financial Markets Using TimeMixer
}
\author{
    Alex Y. LI \\
    Hong Kong Community College \\
    \texttt{23118336A@common.cpce-PolyU.edu.hk}
}
\maketitle

\begin{abstract}
Predicting volatility in financial markets, including stocks, index ETFs, foreign exchange, and cryptocurrencies, remains a challenging task due to the inherent complexity and non-linear dynamics of these time series. In this study, I apply TimeMixer, a state-of-the-art time series forecasting model, to predict the volatility of global financial assets. TimeMixer utilizes a multiscale-mixing approach that effectively captures both short-term and long-term temporal patterns by analyzing data across different scales. My empirical results reveal that while TimeMixer performs exceptionally well in short-term volatility forecasting, its accuracy diminishes for longer-term predictions, particularly in highly volatile markets. These findings highlight TimeMixer’s strength in capturing short-term volatility, making it highly suitable for practical applications in financial risk management, where precise short-term forecasts are critical. However, the model's limitations in long-term forecasting point to potential areas for further refinement.

\vspace{1\baselineskip}
\end{abstract}

\section{INTRODUCTION}

Forecasting volatility in global financial markets, including stocks, index ETFs, foreign exchange, and cryptocurrencies is critical for investors, financial institutions, and policymakers. Successful volatility predictions are essential for risk management, derivative pricing, and portfolio optimization. However, the non-linear behavior and complexity of financial time series, characterized by noise, irregular fluctuations, and abrupt changes, make volatility forecasting a particularly challenging task \cite{1}. Accurately identifying patterns across varying time horizons is difficult due to the interplay between short-term market movements and long-term trends.

Traditional time series forecasting methods, such as  Autoregressive Conditional Heteroskedasticity (ARCH) \cite{2} and Generalized Autoregressive Conditional Heteroskedasticity (GARCH) \cite{3}, as well as machine learning models like  Long Short-Term Memor (LSTM) \cite{4} and Gated Recurrent Units (GRU) \cite{5}, have shown some success. However, they often struggle to fully capture the multiscale dynamics inherent in financial data. While these models may perform well in specific contexts, their limitations become evident when confronted with the diverse temporal dependencies characteristic of financial markets. In particular, short-term fluctuations and long-term trends frequently require distinct treatment, which many existing models fail to adequately address.

To overcome these challenges, this study applies TimeMixer \cite{6}, a state-of-the-art time series forecasting model, to predict the volatility of global financial derivatives. TimeMixer, introduced by Shiyu Wang et al. from Ant Group and Tsinghua University, employs a multiscale-mixing approach to disentangle and analyze both short-term and long-term temporal structures. By leveraging a fully MLP-based architecture, TimeMixer decomposes time series into distinct temporal scales, allowing it to capture both fine-grained seasonal components and broader trend patterns. Its Past-Decomposable-Mixing (PDM) and Future-Multipredictor-Mixing (FMM) blocks enable robust extraction of multiscale information from past data and accurate future predictions.

This multiscale approach is particularly valuable for financial market volatility forecasting, where short-term market reactions can significantly differ from long-term trends. In this study, I leverage TimeMixer’s ability to capture both microscopic and macroscopic temporal dynamics to forecast the volatility of various financial derivatives. The empirical analysis focuses on the model's performance in short-term versus long-term volatility predictions, with findings indicating that TimeMixer is more effective in the short term, where it achieves higher predictive accuracy. This advantage likely stems from TimeMixer's ability to harness fine-scale information crucial for capturing rapid market movements.

While this research does not compare TimeMixer directly with other forecasting models, it serves as an exploratory application of TimeMixer in financial markets. The results demonstrate its potential as a powerful tool for predicting short-term volatility, which is of particular importance to traders and risk managers. Future work will explore how TimeMixer’s multiscale capabilities can be further refined to improve its long-term forecasting accuracy.

\section{RELATED WORK}

\subsection{Traditional Statistical Models for Volatility Forecasting}
Time series forecasting has a long history in financial markets, with traditional statistical models like  Autoregressive Conditional Heteroskedasticity (ARCH) and Generalized Autoregressive Conditional Heteroskedasticity (GARCH) being widely used to model volatility through conditional heteroskedasticity. These models perform well under assumptions of linear relationships and fixed variance but often struggle with the non-linear dynamics and sudden regime shifts typical of financial data. Similarly, models like ARIMA \cite{7} and Vector Autoregression (VAR) \cite{8} are frequently applied to financial time series. While useful for capturing trends and cycles in relatively stationary data, they are limited by their inability to handle non-stationarity and long-term dependencies, both of which are critical for accurately forecasting volatility in more complex markets.

\subsection{Machine Learning and Deep Learning Approaches}
In recent years, machine learning and deep learning models have gained prominence in time series forecasting. Models such as Long Short-Term Memory (LSTM) and Gated Recurrent Units (GRU) are particularly favored in financial forecasting due to their ability to model long-term dependencies. However, these models often require extensive hyperparameter tuning and struggle to capture the multiscale nature of financial time series, where short-term volatility and long-term trends coexist. Similarly, models such as Convolutional Neural Networks (CNNs) \cite{9} and Transformers \cite{10} have been explored for various time series tasks. While Transformers have shown promise in handling longer forecast horizons, they may not be as effective in capturing the short-term fluctuations that are critical for accurate volatility prediction in financial markets, where rapid chan

\subsection{Multiscale Approaches for Time Series Decomposition}
To address the limitations of models that operate on a single scale, multiscale approaches like Wavelet Transforms \cite{11} and Empirical Mode Decomposition (EMD) \cite{12}  have been introduced. These methods aim to capture both fine-grained and coarse-grained patterns by decomposing the time series into multiple scales. While effective in providing a structured analysis of temporal patterns, these techniques often rely on manual decomposition, which can limit their adaptability across diverse datasets and dynamic market conditions. Additionally, these methods are not always capable of automatically learning the most relevant scales, making them less flexible for highly volatile environments like financial markets.

\subsection{TimeMixer: A Novel Multiscale-Mixing Approach}
TimeMixer, introduced by Shiyu Wang et al., represents a novel approach to time series forecasting by automatically learning multiscale representations. Unlike traditional models that require manual decomposition or struggle to balance different temporal dependencies, TimeMixer uses a fully MLP-based \cite{13} architecture with Past-Decomposable-Mixing (PDM) and Future-Multipredictor-Mixing (FMM) blocks to extract and integrate information across multiple time scales. This enables the model to disentangle short-term and long-term dynamics directly from the data. TimeMixer’s architecture is particularly well-suited for financial volatility forecasting, where market behavior often exhibits multiscale patterns. This automatic mixing of temporal scales makes TimeMixer highly effective in capturing both rapid market movements and longer-term trends—two critical components in financial markets that are difficult to model simultaneously using traditional methods.

\subsection{Application of TimeMixer in Financial Markets}
While TimeMixer has been evaluated across a range of time series tasks, its application to financial market volatility prediction remains underexplored, particularly in distinguishing short-term and long-term forecasting accuracy. In this study, I aim to bridge this gap by applying TimeMixer to forecast the volatility of global financial derivatives. Special attention is given to its performance in short-term volatility predictions, which are critical for risk management and trading strategies. By focusing on the model’s ability to capture short-term fluctuations, this research highlights TimeMixer’s potential as a practical tool for financial risk management, where precise and timely forecasts are paramount.
\vspace{1\baselineskip}

Given a series x with one or multiple observed variates, the main objective of time series forecasting
is to utilize past observations (length-P) to obtain the most probable future prediction (length-F). As mentioned earlier, one of the main challenges in accurate forecasting is addressing complex temporal variations. TimeMixer utilizes multiscale mixing to disentangle such variations, enhancing complementary forecasting capabilities across different scales. TimeMixer is built around a multiscale mixing framework, incorporating Past-Decomposable-Mixing for extracting historical information and Future-Multipredictor-Mixing for generating future predictions.
\vspace{1\baselineskip}

\begin{figure}[ht]
    \centering
    \includegraphics[width=16cm,height=5cm]{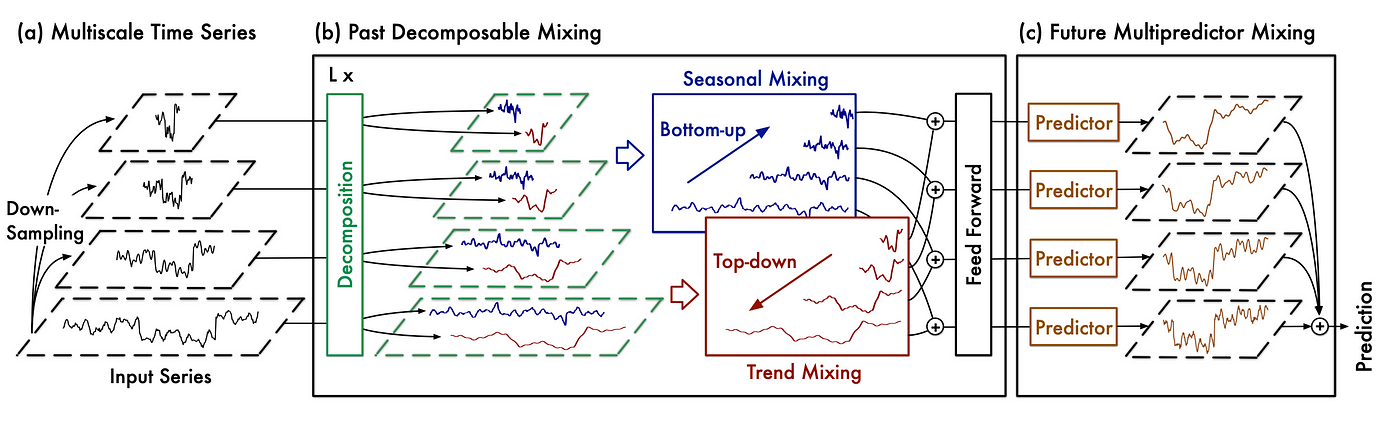}
    \textit{Image source: Shiyu Wang et al., *TimeMixer: Decomposable Multiscale Mixing for Time Series Forecasting*, 2024, ICLR.}
    \caption{Overall architecture of TimeMixer, which consists of (a) Multiscale Time Series, (b) Past Decomposable Mixing and (c) Future Multipredictor Mixing for past observations and future predictions respectively.}
    \label{fig:TimeMixer_1}
    
\end{figure}

\subsection{MULTISCALE MIXING ARCHITECTURE}
The core idea behind TimeMixer is its multiscale mixing architecture. The model divides the input time series into multiple time scales, including short-term, medium-term, and long-term windows. Each time scale captures patterns at different temporal resolutions. By mixing these scales, TimeMixer can learn from both short-term variations and long-term trends, creating a more comprehensive understanding of the data. This multiscale approach helps the model predict future values more accurately.

\[
X = \{ x_0, \ldots, x_M \}, \quad \text{where } x_m \in \mathbb{R}^{\left\lfloor \frac{P}{2^m} \right\rfloor \times C}, \quad m \in \{ 0, \ldots, M \}, \, C
\]

\subsection{Past-Decomposable-Mixing (PDM)}
The Past-Decomposable-Mixing (PDM) mechanism is used to extract useful information from past data. TimeMixer breaks down the past time series into different components, each representing a unique time scale. This enables the model to focus on significant historical trends and seasonal patterns while disregarding irrelevant or noisy short-term fluctuations. By decomposing the past data in this manner, TimeMixer can identify meaningful patterns that are crucial for future predictions.

\begin{figure}[ht]
    \centering
    \includegraphics[width=16cm,height=4cm]{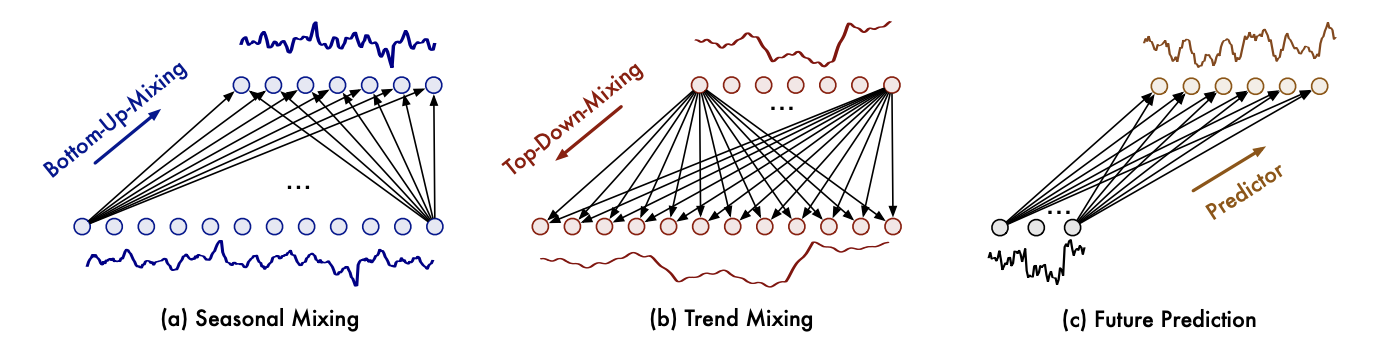}
        \textit{Image source: Shiyu Wang et al., *TimeMixer: Decomposable Multiscale Mixing for Time Series Forecasting*, 2024, ICLR.}
    \caption{The temporal linear layer in seasonal mixing (a), trend mixing (b) and future prediction (c).}
    \label{fig:TimeMixer_2}
\end{figure}

\[
s^{l}_{m}, \quad t^{l}_{m} = \text{SeriesDecomp}(x^{l}_{m}), \quad m \in \{0, \ldots, M\},
\]
\[
X^{l} = X^{l-1} + \text{FeedForward} \left(  \text{S-Mix} \left( \left\{ s^{l}_{m} \right\}_{m=0}^{M} \right) + \text{T-Mix} \left( \left\{ t^{l}_{m} \right\}_{m=0}^{M} \right) \right),
\]

Seasonal Mixing

\[
\text{for } m: 1 \to M \text{ do: } s^{l}_{m} = s^{l}_{m} + \text{Bottom-Up-Mixing}(s^{l}_{m-1})
\]

Trend Mixing

\[
\text{for } m: (M - 1) \to 0 \text{ do: } t^{l}_{m} = t^{l}_{m} + \text{Top-Down-Mixing}(t^{l}_{m+1})
\]

\subsection{Future-Multipredictor-Mixing (FMM)}
The Future-Multipredictor-Mixing (FMM) mechanism is responsible for making predictions about future values. TimeMixer generates multiple predictions for future time steps, with each prediction focusing on a different time scale. These predictions are then mixed together to form a final, more accurate forecast. By using multiple predictors, TimeMixer can account for uncertainties and variations in the data, improving the reliability of its forecasts.

\[
\hat{x}_m = \text{Predictor}_m(x^{L}_{m}), \quad m \in \{0, \ldots, M\}, \quad \hat{x}_m = \sum_{m=0}^{M} \hat{x}_m
\]

\section{EXPERIMENTS}

\subsection{Experiment Overview}
This paper presents an empirical study on predicting volatility in financial markets—including stocks, index ETFs, foreign exchange, and cryptocurrencies—using the TimeMixer model. The study leverages OHLCV (Open, High, Low, Close, Volume) data obtained from Yahoo Finance to predict volatility. The validation set comprises 10\% of the training set size. The results demonstrate that TimeMixer excels in predicting short-term volatility, outperforming its long-term forecasts. By utilizing historical market data, the model shows a significant advantage in forecasting near-term volatility. TimeMixer's multiscale-mixing architecture allows it to effectively capture short-term temporal dependencies, making it particularly well-suited for short-horizon predictions. However, its predictive accuracy for longer-term trends is comparatively lower.

For each asset, the dataset includes daily OHLCV values spanning multiple years. The volatility for each asset is calculated using the following formula:

\[
\sigma = \sqrt{252} \times \text{std}\left( \log \frac{P_t}{P_{t-1}}, \text{window} = 21 \right)
\]

Where:
\begin{itemize}
    \item \(\log \frac{P_t}{P_{t-1}}\) represents the daily log returns.
    \item \(\text{std}()\) denotes the standard deviation.
    \item \(\text{window} = 21\) refers to the rolling window of 21 days (approximately one trading month).
    \item \(\sqrt{252}\) annualizes the volatility, assuming 252 trading days per year.
\end{itemize}

The following table summarizes the experimental data across various datasets, detailing the ticker symbols, date ranges, and corresponding asset names. The ticker symbol serves as a unique identifier for each asset, which may represent stocks, bonds, commodities, or other financial instruments. The date range indicates the period during which the data was collected or analyzed. The asset name refers to the full legal name or formal description of the entity associated with each ticker. This dataset provides a comprehensive overview of the performance of different asset types across specific timeframes, offering valuable insights into their behavior within the studied periods.

\subsection{Stock Dataset}

\begin{table}[H]
\centering
\label{table:Stcok_1}
\begin{tabularx}{\textwidth}{l|X|X}

\hline
\textbf{Ticker Symbol}   & \textbf{Date Range} & \textbf{Company} \\ \hline
AAPL   & \hspace{1pt}2010 - 2023\hspace{1pt} & Apple Inc. \\ \hdashline
GOOGL  & \hspace{1pt}2010 - 2023\hspace{1pt} & Alphabet Inc. \\ \hdashline
MSFT   & \hspace{1pt}2010 - 2023\hspace{1pt} & Microsoft Corporation \\ \hdashline
AMZN   & \hspace{1pt}2010 - 2023\hspace{1pt} & Amazon.com, Inc. \\ \hdashline
TSLA   & \hspace{1pt}2010 - 2023\hspace{1pt} & Tesla, Inc. \\ \hdashline
BRK-B  & \hspace{1pt}2010 - 2023\hspace{1pt} & Berkshire Hathaway Inc. \\ \hdashline
NVDA   & \hspace{1pt}2010 - 2023\hspace{1pt} & NVIDIA Corporation \\ \hdashline
META   & \hspace{1pt}2012 - 2023\hspace{1pt} & Meta Platforms, Inc. \\ \hdashline
V      & \hspace{1pt}2010 - 2023\hspace{1pt} & Visa Inc. \\ \hdashline
JNJ    & \hspace{1pt}2010 - 2023\hspace{1pt} & Johnson \& Johnson \\ \hline
\end{tabularx}
\caption{Stock Datasets: Including Ticker Symbols, Date Ranges and Company.}
\end{table}
\subsection{Index ETF Dataset}

\begin{table}[H]
\centering
\label{table:Index_1}
\begin{tabularx}{\textwidth}{l|X|X}

\hline
\textbf{Ticker Symbol}   & \textbf{Date Range} & \textbf{ETF} \\ \hline
SPY   & \hspace{1pt}2010 - 2023\hspace{1pt} & SPDR S\&P 500 ETF Trust \\ \hdashline
QQQ  & \hspace{1pt}2010 - 2023\hspace{1pt} & Invesco QQQ Trust \\ \hdashline
VTI   & \hspace{1pt}2010 - 2023\hspace{1pt} & Vanguard Total Stock Market Index Fund ETF Shares \\ \hdashline
EEM   & \hspace{1pt}2010 - 2023\hspace{1pt} & iShares MSCI Emerging Markets ETF \\ \hdashline
EFA   & \hspace{1pt}2010 - 2023\hspace{1pt} & iShares MSCI EAFE ETF \\ \hdashline
VWO  & \hspace{1pt}2010 - 2023\hspace{1pt} & Vanguard Emerging Markets Stock Index Fund \\ \hdashline
IWM   & \hspace{1pt}2010 - 2023\hspace{1pt} & iShares Russell 2000 ETF \\ \hdashline
GLD   & \hspace{1pt}2010 - 2023\hspace{1pt} & SPDR Gold Shares \\ \hdashline
GOVT      & \hspace{1pt}2012 - 2023\hspace{1pt} & iShares US Treasury Bond ETF \\ \hdashline
SCHD    & \hspace{1pt}2011 - 2023\hspace{1pt} & Schwab U.S. Dividend Equity ETF \\ \hline
\end{tabularx}
\caption{Index ETF Data Sets: Including Ticker Symbols, Date Ranges and ETF.}
\end{table}
\subsection{Forex Dataset}

\begin{table}[H]
\centering
\label{table:Forex_1}
\begin{tabularx}{\textwidth}{l|X|X}

\hline
\textbf{Ticker Symbol}   & \textbf{Date Range} & \textbf{Currency} \\ \hline
EURUSD   & \hspace{5pt}2010 - 2023\hspace{5pt} & EUR/USD \\ \hdashline
USDJPY  & \hspace{5pt}2010 - 2023\hspace{5pt} & USD/JPY \\ \hdashline
GBPUSD   & \hspace{5pt}2010 - 2023\hspace{5pt} & GBP/USD \\ \hdashline
AUDUSD   & \hspace{5pt}2010 - 2023\hspace{5pt} & AUD/USD \\ \hdashline
USDCAD   & \hspace{5pt}2010 - 2023\hspace{5pt} & USD/CAD \\ \hdashline
USDCHF  & \hspace{5pt}2010 - 2023\hspace{5pt} & USD/CHF \\ \hdashline
EURGBP   & \hspace{5pt}2010 - 2023\hspace{5pt} & EUR/GBP \\ \hdashline
EURJPY   & \hspace{5pt}2010 - 2023\hspace{5pt} & EUR/JPY \\ \hdashline
GBPJPY      & \hspace{5pt}2010 - 2023\hspace{5pt} & GBP/JPY \\ \hdashline
AUDJPY   & \hspace{5pt}2010 - 2023\hspace{5pt} & AUD/JPY \\ \hline
\end{tabularx}
\caption{Forex Datasets: Including Ticker Symbols, Date Ranges and Currency.}
\end{table}
\subsection{Cryptocurrency Dataset}

\begin{table}[H]
\centering
\label{table:Cryptocurrency_1}
\begin{tabularx}{\textwidth}{l|X|X}

\hline
\textbf{Ticker Symbol}   & \textbf{Date Range} & \textbf{Cryptocurrency} \\ \hline
BTCUSD   & \hspace{5pt}2014 - 2023\hspace{5pt} & Bitcoin/USD \\ \hdashline
ETHUSD  & \hspace{5pt}2017 - 2023\hspace{5pt} & Ethereum/USD \\ \hdashline
LTCUSD   & \hspace{5pt}2014 - 2023\hspace{5pt} & Litecoin/USD \\ \hdashline
BCHUSD   & \hspace{5pt}2017 - 2023\hspace{5pt} & Bitcoin Cash/USD \\ \hdashline
DOGEUSD   & \hspace{5pt}2017 - 2023\hspace{5pt} & Dogecoin/USD \\ \hdashline
XRPUSD  & \hspace{5pt}2017 - 2023\hspace{5pt} & XRP/USD \\ \hdashline
ADAUSD   & \hspace{5pt}2017 - 2023\hspace{5pt} & Cardano/USD \\ \hdashline
DOTUSD   & \hspace{5pt}2020 - 2023\hspace{5pt} & Polkadot/USD \\ \hdashline
BNBUSD      & \hspace{5pt}2017 - 2023\hspace{5pt} & BNB/USD \\ \hdashline
SOLUSD    & \hspace{5pt}2020 - 2023\hspace{5pt} & Solana/USD \\ \hline
\end{tabularx}
\caption{Cryptocurrency Datasets: Including Ticker Symbols, Date Ranges and Cryptocurrency.}
\end{table}

\newpage

\section{MAIN RESULTS}

\subsection{Stock Volatility Forecasting Results}

\begin{table}[H]
\centering
\renewcommand{\arraystretch}{1.0} 
\resizebox{\textwidth}{!}{
\begin{tabular}{l|c|cccccccccc}
\toprule
\textbf{} & \textbf{} & \textbf{AAPL} & \textbf{GOOGL} & \textbf{MSFT} & \textbf{AMZN} & \textbf{TSLA} & \textbf{BRK-B} & \textbf{NVDA} & \textbf{META} & \textbf{V} & \textbf{JNJ} \\
\midrule
\multirow{3}{*}{\textbf{12 Days}} 
 & \textbf{MAE}   & 0.0037 & 0.0159 & 0.0313 & 0.0114 & 0.0170 & 0.0143 & 0.0109 & 0.0190 & 0.0053 & 0.0139 \\
 & \textbf{MSE}   & 0.0001 & 0.0003 & 0.0011 & 0.0002 & 0.0004 & 0.0002 & 0.0002 & 0.0005 & 0.0001 & 0.0003 \\
 & \textbf{RMSE}  & 0.0059 & 0.0170 & 0.0333 & 0.0147 & 0.0192 & 0.0156 & 0.0134 & 0.0230 & 0.0064 & 0.0159 \\
\midrule
\multirow{3}{*}{\textbf{96 Days}} 
 & \textbf{MAE}   & 0.0354 & 0.0655 & 0.0245 & 0.0654 & 0.0898 & 0.0149 & 0.0820 & 0.0275 & 0.0362 & 0.0321 \\
 & \textbf{MSE}   & 0.0018 & 0.0074 & 0.0010 & 0.0057 & 0.0112 & 0.0003 & 0.0089 & 0.0011 & 0.0021 & 0.0013 \\
 & \textbf{RMSE}  & 0.0419 & 0.0858 & 0.0309 & 0.0753 & 0.1058 & 0.0182 & 0.0945 & 0.0330 & 0.0454 & 0.0360 \\
\midrule
\multirow{3}{*}{\textbf{192 Days}} 
 & \textbf{MAE}   & 0.0308 & 0.0530 & 0.0839 & 0.0517 & 0.0805 & 0.0348 & 0.1254 & 0.0634 & 0.0499 & 0.0456 \\
 & \textbf{MSE}   & 0.0015 & 0.0052 & 0.0082 & 0.0042 & 0.0098 & 0.0017 & 0.0201 & 0.0060 & 0.0034 & 0.0038 \\
 & \textbf{RMSE}  & 0.0392 & 0.0720 & 0.0903 & 0.0648 & 0.0988 & 0.0411 & 0.1417 & 0.0777 & 0.0579 & 0.0614 \\
\midrule
\multirow{3}{*}{\textbf{336 Days}} 
 & \textbf{MAE}   & 0.1446 & 0.0973 & 0.1022 & 0.1167 & 0.2168 & 0.0645 & 0.2883 & 0.1334 & 0.0422 & 0.0367 \\
 & \textbf{MSE}   & 0.0236 & 0.0136 & 0.0136 & 0.0180 & 0.0555 & 0.0053 & 0.1058 & 0.0421 & 0.0026 & 0.0024 \\
 & \textbf{RMSE}  & 0.1536 & 0.1168 & 0.1166 & 0.1341 & 0.2356 & 0.0728 & 0.3253 & 0.2052 & 0.0513 & 0.0486 \\
\midrule
\multirow{3}{*}{\textbf{720 Days}} 
 & \textbf{MAE}   & 0.0837 & 0.1094 & 0.0983 & 0.1858 & 0.1886 & 0.0543 & 0.1460 & 0.1730 & 0.0768 & 0.0440 \\
 & \textbf{MSE}   & 0.0125 & 0.0192 & 0.0148 & 0.0496 & 0.0571 & 0.0048 & 0.0290 & 0.0661 & 0.0111 & 0.0032 \\
 & \textbf{RMSE}  & 0.1120 & 0.1387 & 0.1217 & 0.2226 & 0.2390 & 0.0693 & 0.1702 & 0.2570 & 0.1053 & 0.0564 \\
\bottomrule
\end{tabular}
}
\caption{Stock Volatility Forecasting Results: Mean Absolute Error (MAE), Mean Squared Error (MSE), and Root Mean Squared Error (RMSE) for Different Prediction Lengths {12, 96, 192, 336, 720}.}
\label{table:Stock_2}
\end{table}

Based on the data, the model performs well in short-term forecasting  (12 days), particularly for less volatile stocks like AAPL and V. For instance, AAPL shows a low MAE of 0.0037 and RMSE of 0.0059 over a 12-day period, while V has a similarly low MAE of 0.0053 and RMSE of 0.0064, indicating the model's ability to capture short-term market fluctuations with high accuracy. Even for more volatile stocks such as TSLA and META, the model maintains reasonable error rates, with TSLA showing an MAE of 0.0170 and RMSE of 0.0192 over the same period, demonstrating the model's robustness in short-term predictions.

However, as the forecasting horizon extends, the model's accuracy decreases significantly, particularly for highly volatile stocks. For example, NVDA's MAE increases to 0.2883 at 336 days and remains relatively high at 0.1460 for 720 days. This suggests that the model struggles to maintain accuracy in long-term forecasts for stocks with high volatility. In contrast, more stable stocks like BRK-B and JNJ show comparatively lower errors over longer periods, with BRK-B having an MAE of 0.0645 at 336 days and 0.0543 at 720 days. This indicates that the model is better suited for long-term forecasting of less volatile stocks.

\subsection{Index ETF Volatility Forecasting Results}

\begin{table}[H]
\centering
\renewcommand{\arraystretch}{1.0} 
\resizebox{\textwidth}{!}{
\begin{tabular}{l|c|cccccccccc}
\toprule
\textbf{} & \textbf{} & \textbf{SPY} & \textbf{QQQ} & \textbf{VTI} & \textbf{EEM} & \textbf{EFA} & \textbf{VWO} & \textbf{IWM} & \textbf{GLD} & \textbf{GOVT} & \textbf{SCHD} \\
\midrule
\multirow{3}{*}{\textbf{12 Days}} 
 & \textbf{MAE}   & 0.0151 & 0.0071 & 0.0154 & 0.0195 & 0.0049 & 0.0205 & 0.0062 & 0.0047 & 0.0022 & 0.0080 \\
 & \textbf{MSE}   & 0.0003 & 0.0001 & 0.0003 & 0.0005 & 0.0001 & 0.0005 & 0.0001 & 0.0001 & 0.0001 & 0.0001 \\
 & \textbf{RMSE}  & 0.0167 & 0.0080 & 0.0169 & 0.0212 & 0.0056 & 0.0214 & 0.0075 & 0.0052 & 0.0028 & 0.0090 \\
\midrule
\multirow{3}{*}{\textbf{96 Days}} 
 & \textbf{MAE}   & 0.0193 & 0.0288 & 0.0222 & 0.0155 & 0.0158 & 0.0123 & 0.0696 & 0.0654 & 0.0168 & 0.0226 \\
 & \textbf{MSE}   & 0.0006 & 0.0012 & 0.0007 & 0.0003 & 0.0003 & 0.0003 & 0.0067 & 0.0051 & 0.0003 & 0.0010 \\
 & \textbf{RMSE}  & 0.0240 & 0.0343 & 0.0272 & 0.0183 & 0.0181 & 0.0168 & 0.0817 & 0.0715 & 0.0181 & 0.0311 \\
\midrule
\multirow{3}{*}{\textbf{192 Days}} 
 & \textbf{MAE}   & 0.0195 & 0.0191 & 0.0221 & 0.0154 & 0.0140 & 0.0151 & 0.0390 & 0.0323 & 0.0086 & 0.0399 \\
 & \textbf{MSE}   & 0.0006 & 0.0007 & 0.0006 & 0.0004 & 0.0003 & 0.0003 & 0.0022 & 0.0013 & 0.0001 & 0.0022 \\
 & \textbf{RMSE}  & 0.0235 & 0.0264 & 0.0246 & 0.0188 & 0.0167 & 0.0185 & 0.0471 & 0.0359 & 0.0116 & 0.0464 \\
\midrule
\multirow{3}{*}{\textbf{336 Days}} 
 & \textbf{MAE}   & 0.0679 & 0.0595 & 0.0444 & 0.0295 & 0.0448 & 0.0187 & 0.0457 & 0.0302 & 0.0361 & 0.0350 \\
 & \textbf{MSE}   & 0.0057 & 0.0044 & 0.0025 & 0.0013 & 0.0025 & 0.0007 & 0.0034 & 0.0014 & 0.0020 & 0.0016 \\
 & \textbf{RMSE}  & 0.0752 & 0.0666 & 0.0498 & 0.0363 & 0.0500 & 0.0268 & 0.0580 & 0.0374 & 0.0444 & 0.0399 \\
\midrule
\multirow{3}{*}{\textbf{720 Days}} 
 & \textbf{MAE}   & 0.0669 & 0.0825 & 0.0593 & 0.0468 & 0.0876 & 0.0426 & 0.0562 & 0.0586 & 0.0226 & 0.0544 \\
 & \textbf{MSE}   & 0.0077 & 0.0106 & 0.0064 & 0.0044 & 0.0108 & 0.0041 & 0.0049 & 0.0045 & 0.0009 & 0.0043 \\
 & \textbf{RMSE}  & 0.0876 & 0.1029 & 0.0798 & 0.0663 & 0.1041 & 0.0639 & 0.0699 & 0.0674 & 0.0303 & 0.0659 \\
\bottomrule
\end{tabular}
}
\caption{Index ETF Volatility Forecasting Results: Mean Absolute Error (MAE), Mean Squared Error (MSE), and Root Mean Squared Error (RMSE) for Different Prediction Lengths {12, 96, 192, 336, 720}.}
\label{table:Intex_2}
\end{table}

The model demonstrates strong performance in short-term forecasts (12 days), especially for relatively stable assets like GLD and GOVT, where the MAE values are as low as 0.0047 and 0.0022, respectively, and the RMSE values are 0.0052 for GLD and 0.0028 for GOVT. Even for more volatile assets, such as QQQ and IWM, the errors remain manageable, with QQQ having an MAE of 0.0071 and RMSE of 0.0080, and IWM showing an MAE of 0.0062 and RMSE of 0.0075. This suggests the model effectively captures short-term volatility patterns across a range of assets.

However, as the prediction horizon extends, the model's accuracy diminishes, particularly for more volatile ETFs like QQQ, IWM, and EEM. For example, QQQ's MAE rises to 0.0595 at 336 days and 0.0825 at 720 days, while IWM's MAE increases to 0.0457 at 336 days and 0.0562 at 720 days. Similarly, EEM's RMSE grows from 0.0212 at 12 days to 0.0663 at 720 days. These increases in error metrics highlight the model's limitations in capturing long-term volatility dynamics, where macroeconomic factors and market unpredictability introduce greater complexity.

On the other hand, the model maintains relatively better long-term performance for more stable assets. For example, GOVT's MAE remains low at 0.0361 over 336 days and 0.0226 over 720 days, while GLD's RMSE is 0.0374 at 336 days and 0.0674 at 720 days. This indicates that the model is more reliable for long-term forecasts of low-volatility assets, such as government bonds and gold, compared to more volatile equity-based ETFs.

\subsection{Forex Volatility Forecasting Results}

\begin{table}[H]
\centering
\renewcommand{\arraystretch}{1.0} 
\resizebox{\textwidth}{!}{
\begin{tabular}{l|c|cccccccccc}
\toprule
\textbf{} & \textbf{} & \textbf{EURUSD} & \textbf{USDJPY} & \textbf{GBPUSD} & \textbf{AUDUSD} & \textbf{USDCAD} & \textbf{USDCHF} & \textbf{EURGBP} & \textbf{EURJPY} & \textbf{GBPJPY} & \textbf{AUDJPY} \\
\midrule
\multirow{3}{*}{\textbf{12 Days}} 
& \textbf{MAE}  & 0.0097 & 0.0022 & 0.0100 & 0.0038 & 0.0015 & 0.0094 & 0.0032 & 0.0128 & 0.0109 & 0.0120 \\
& \textbf{MSE}  & 0.0001 & 0.0001 & 0.0001 & 0.0001 & 0.0001 & 0.0001 & 0.0001 & 0.0002 & 0.0001 & 0.0002 \\
& \textbf{RMSE} & 0.0098 & 0.0027 & 0.0102 & 0.0040 & 0.0019 & 0.0109 & 0.0033 & 0.0137 & 0.0122 & 0.0126 \\
\midrule
\multirow{3}{*}{\textbf{96 Days}} 
& \textbf{MAE}  & 0.0083 & 0.0225 & 0.0182 & 0.0277 & 0.0077 & 0.0080 & 0.0057 & 0.0132 & 0.0198 & 0.0109 \\
& \textbf{MSE}  & 0.0001 & 0.0008 & 0.0005 & 0.0009 & 0.0001 & 0.0001 & 0.0001 & 0.0003 & 0.0007 & 0.0002 \\
& \textbf{RMSE} & 0.0101 & 0.0289 & 0.0215 & 0.0298 & 0.0090 & 0.0094 & 0.0065 & 0.0163 & 0.0260 & 0.0127 \\
\midrule
\multirow{3}{*}{\textbf{192 Days}} 
& \textbf{MAE}  & 0.0071 & 0.0184 & 0.0114 & 0.0117 & 0.0077 & 0.0218 & 0.0071 & 0.0193 & 0.0249 & 0.0155 \\
& \textbf{MSE}  & 0.0001 & 0.0005 & 0.0002 & 0.0002 & 0.0001 & 0.0006 & 0.0001 & 0.0005 & 0.0009 & 0.0004 \\
& \textbf{RMSE} & 0.0083 & 0.0226 & 0.0148 & 0.0144 & 0.0097 & 0.0246 & 0.0084 & 0.0219 & 0.0299 & 0.0200 \\
\midrule
\multirow{3}{*}{\textbf{336 Days}} 
& \textbf{MAE}  & 0.0281 & 0.0330 & 0.0496 & 0.0302 & 0.0350 & 0.0255 & 0.0666 & 0.0192 & 0.0257 & 0.0382 \\
& \textbf{MSE}  & 0.0010 & 0.0019 & 0.0027 & 0.0011 & 0.0014 & 0.0009 & 0.0052 & 0.0006 & 0.0009 & 0.0021 \\
& \textbf{RMSE} & 0.0316 & 0.0434 & 0.0524 & 0.0337 & 0.0371 & 0.0292 & 0.0720 & 0.0239 & 0.0306 & 0.0460 \\
\midrule
\multirow{3}{*}{\textbf{720 Days}} 
& \textbf{MAE}  & 0.0229 & 0.0557 & 0.0403 & 0.0367 & 0.0174 & 0.0255 & 0.0362 & 0.0219 & 0.0422 & 0.0275 \\
& \textbf{MSE}  & 0.0008 & 0.0042 & 0.0032 & 0.0021 & 0.0005 & 0.0011 & 0.0110 & 0.0009 & 0.0028 & 0.0013 \\
& \textbf{RMSE} & 0.0289 & 0.0646 & 0.0562 & 0.0460 & 0.0232 & 0.0332 & 0.1049 & 0.0292 & 0.0528 & 0.0363 \\
\bottomrule
\end{tabular}
}
\caption{Forex Volatility Forecasting Results: Mean Absolute Error (MAE), Mean Squared Error (MSE), and Root Mean Squared Error (RMSE) for Different Prediction Lengths {12, 96, 192, 336, 720}.}
\label{table:Forex_2}
\end{table}

The model shows strong performance in short-term forecasting (12 days), particularly for relatively stable currency pairs such as USD/JPY and USD/CAD, which achieve the lowest MAE values of 0.0022 and 0.0015, respectively. These pairs also exhibit low RMSE values, with USD/JPY at 0.0027 and USD/CAD at 0.0019, indicating the model's high accuracy in predicting short-term exchange rate movements. Major currency pairs like EUR/USD and GBP/USD also perform reasonably well, with moderate MAE values of 0.0097 and 0.0100, respectively. Even more volatile pairs such as EUR/JPY and GBP/JPY maintain RMSE values under 0.0140, showcasing the model's ability to handle short-term volatility effectively.

However, the model's accuracy declines as the forecasting horizon extends to 96 days, particularly for more volatile pairs like AUD/USD and USD/JPY. For example, AUD/USD's MAE rises significantly to 0.0277, and USD/JPY's MAE increases to 0.0225, demonstrating the model's challenges in maintaining precision over medium-term periods. Conversely, USD/CAD continues to exhibit relatively low errors with an MAE of 0.0077, suggesting that the model is better suited for forecasting lower-volatility currency pairs over both short and medium terms.

The model's performance deteriorates further in the long-term (192 days and beyond), especially for highly volatile currency pairs like GBP/JPY, EUR/GBP, and EUR/JPY. At the 336-day horizon, EUR/GBP's MAE jumps to 0.0666, and GBP/JPY reaches 0.0382, reflecting the model's limitations in handling long-term volatility for these pairs. AUD/JPY also faces challenges, with an RMSE of 0.0460, indicating significant difficulty in predicting its exchange rate movements over an extended period. EUR/GBP stands out with the highest RMSE of 0.0720 among all pairs, suggesting that the model struggles the most with this cross-currency pair in long-term forecasts.

By the 720-day horizon, the model's limitations become even more apparent for pairs like GBP/JPY and EUR/GBP, whose RMSE values rise to 0.0528 and 0.1049, respectively. This highlights the increasing difficulty the model faces in predicting long-term volatility for cross-currency and highly volatile pairs. Despite these challenges, the model performs comparatively better for more stable pairs like USD/CHF and USD/CAD, which maintain lower MAE values of 0.0255 and 0.0174, respectively, even at this extended horizon. This suggests that the model may be more reliable for long-term forecasting of lower-volatility pairs, whereas improvements are needed to enhance its performance for volatile and cross-currency pairs over longer horizons.

\subsection{Cryptocurrency Volatility Results}

\begin{table}[H]
\centering
\renewcommand{\arraystretch}{1.0} 
\resizebox{\textwidth}{!}{
\begin{tabular}{l|c|cccccccccc}
\toprule
\textbf{} & \textbf{} & \textbf{BTCUSD} & \textbf{ETHUSD} & \textbf{LTCUSD} & \textbf{BCHUSD} & \textbf{DOGEUSD} & \textbf{XRPUSD} & \textbf{ADAUSD} & \textbf{DOTUSD} & \textbf{BNBUSD} & \textbf{SOL-USD} \\ 
\midrule
\multirow{3}{*}{\textbf{12 Days}} 
& \textbf{MAE}  & 0.1249 & 0.0146 & 0.0121 & 0.0875 & 0.0211 & 0.2060 & 0.0294 & 0.1540 & 0.2344 & 0.1230 \\
& \textbf{MSE}  & 0.0227 & 0.0003 & 0.0003 & 0.0146 & 0.0006 & 0.0635 & 0.0017 & 0.0282 & 0.0783 & 0.0167 \\
& \textbf{RMSE} & 0.1508 & 0.0174 & 0.0167 & 0.1210 & 0.0249 & 0.2521 & 0.0414 & 0.1680 & 0.2798 & 0.1294 \\
\midrule
\multirow{3}{*}{\textbf{96 Days}} 
& \textbf{MAE}  & 0.0583 & 0.1069 & 0.0545 & 0.1509 & 0.2789 & 0.2826 & 0.2623 & 0.2501 & 0.1264 & 0.3019 \\
& \textbf{MSE}  & 0.0056 & 0.0182 & 0.0043 & 0.0403 & 0.1146 & 0.3144 & 0.1286 & 0.0966 & 0.0290 & 0.1473 \\
& \textbf{RMSE} & 0.0750 & 0.1350 & 0.0656 & 0.2008 & 0.3385 & 0.5607 & 0.3586 & 0.3109 & 0.1702 & 0.3839 \\
\midrule
\multirow{3}{*}{\textbf{192 Days}} 
& \textbf{MAE}  & 0.1334 & 0.1039 & 0.1942 & 0.2913 & 0.1625 & 0.3369 & 0.2228 & 0.1632 & 0.2164 & 0.2573 \\
& \textbf{MSE}  & 0.0290 & 0.0170 & 0.0526 & 0.1793 & 0.0430 & 0.2838 & 0.0879 & 0.0444 & 0.0576 & 0.0995 \\
& \textbf{RMSE} & 0.1703 & 0.1303 & 0.2294 & 0.4234 & 0.2073 & 0.5327 & 0.2966 & 0.2106 & 0.2399 & 0.3155 \\
\midrule
\multirow{3}{*}{\textbf{336 Days}} 
& \textbf{MAE}  & 0.2135 & 0.2471 & 0.1365 & 0.2767 & 0.3259 & 0.2577 & 0.2647 & 0.3287 & 0.1823 & 0.7147 \\
& \textbf{MSE}  & 0.0746 & 0.1012 & 0.0394 & 0.1702 & 0.1682 & 0.1293 & 0.1032 & 0.1623 & 0.0485 & 0.7923 \\
& \textbf{RMSE} & 0.2732 & 0.3182 & 0.1985 & 0.4125 & 0.4101 & 0.3596 & 0.3212 & 0.4028 & 0.2201 & 0.8901 \\
\midrule
\multirow{3}{*}{\textbf{720 Days}} 
& \textbf{MAE}  & 0.1694 & 0.2215 & 0.2267 & 0.2595 & 0.3713 & 0.0341 & 0.2603 & 0.4271 & 0.2111 & 0.2591 \\
& \textbf{MSE}  & 0.0388 & 0.0804 & 0.0715 & 0.1155 & 0.2343 & 0.0014 & 0.1230 & 0.2408 & 0.0636 & 0.1550 \\
& \textbf{RMSE} & 0.1969 & 0.2836 & 0.2674 & 0.3398 & 0.4840 & 0.0379 & 0.3506 & 0.4907 & 0.2523 & 0.3938 \\
\bottomrule
\end{tabular}
}

\caption{Cryptocurrency Volatility Forecasting Results: Mean Absolute Error (MAE), Mean Squared Error (MSE), and Root Mean Squared Error (RMSE) for Different Prediction Lengths {12, 96, 192, 336, 720}.}
\label{table:Cryptocurrency_2}
\end{table}

The model demonstrates strong short-term forecasting (12 days) performance for certain cryptocurrencies, particularly for ETH/USD and LTC/USD, which exhibit relatively low error metrics. For example, ETH/USD achieves an MAE of 0.0146 and an RMSE of 0.0174, while LTC/USD shows an MAE of 0.0121 and RMSE of 0.0167, indicating that the model is able to capture short-term price movements effectively for these assets. Even for more volatile assets like BTC/USD, the model maintains reasonable accuracy with an MAE of 0.1249 and RMSE of 0.1508, which is notable given Bitcoin's historical volatility. However, the model struggles more with highly volatile cryptocurrencies such as XRP/USD and BNB/USD, where the RMSE reaches 0.2521 and 0.2798, respectively, suggesting that the model may have difficulty managing short-term price fluctuations for these more unpredictable assets.

As the prediction horizon extends to 96 days, the model's accuracy declines, particularly for highly volatile assets like DOGE/USD and ADA/USD. For instance, DOGE/USD sees a significant increase in MAE to 0.2789 and RMSE to 0.3385, while ADA/USD reaches an MAE of 0.2623 and RMSE of 0.3586. This reflects the challenges the model faces in maintaining accuracy over medium-term periods, where volatility is more difficult to predict. In contrast, BTC/USD continues to perform relatively well with an MAE of 0.0583 and RMSE of 0.0750, suggesting that the model is better suited for forecasting more established and less volatile cryptocurrencies compared to newer or more speculative assets like DOT/USD or ADA/USD.

The model's performance deteriorates further for long-term forecasts, especially for highly volatile assets like SOL/USD and DOGE/USD. At the 336-day horizon, SOL/USD experiences a sharp increase in MAE to 0.7147 and RMSE to 0.8901, indicating that the model struggles significantly to predict long-term price movements for this asset. Similarly, DOGE/USD shows substantial forecasting errors, with an RMSE of 0.4101, reflecting the model's limitations in handling long-term volatility for highly speculative cryptocurrencies.

At the 720-day horizon, the model's limitations become even more evident for assets like DOGE/USD and ADA/USD, with RMSE values of 0.4840 and 0.4907, respectively. This suggests that the model faces increasing difficulty in predicting long-term price trends for these more volatile cryptocurrencies. On the other hand, BTC/USD and ETH/USD still perform comparatively better, with BTC/USD maintaining an RMSE of 0.1969 and ETH/USD at 0.2836, indicating that the model is better suited for forecasting more established assets with longer market histories. However, even for these relatively stable assets, the rising error margins underscore the challenges of long-term forecasting in the highly volatile cryptocurrency market.


\section{CONCLUSION}
The model exhibits robust forecasting capabilities, particularly in short-term predictions for relatively stable and low-volatility assets across a range of categories, including stocks, ETFs, foreign exchange, and cryptocurrencies. It performs well in predicting short-term price movements for assets like BTC/USD, ETH/USD, and AAPL, demonstrating its strength in handling more established and less volatile assets. However, the model struggles with highly volatile assets such as DOGE/USD, SOL/USD, and TSLA, where error metrics increase notably. As the prediction horizon extends, the model's accuracy declines, particularly for highly volatile assets, revealing its limitations in managing long-term forecasts.

\subsection{Limitations}
The primary limitation of the model is its diminishing accuracy over extended forecasting horizons, especially for volatile assets. The model struggles significantly with long-term volatility, as observed in assets like DOGE/USD and AUD/USD, where the Mean Absolute Error (MAE) and Root Mean Squared Error (RMSE) values rise considerably over time. This suggests that the model is less effective at capturing the complex and unpredictable price dynamics of highly volatile assets in long-term forecasts.

\subsection{Future Work}
To enhance the model’s performance in long-term forecasting, future work should explore the integration of macroeconomic indicators (e.g., interest rates, inflation, and GDP growth) and sentiment analysis from news and social media to better capture the broader market environment. Incorporating these elements could enable the model to better predict market shifts, especially for volatile assets. Additionally, developing customized models for specific asset classes, such as stocks, foreign exchange, and cryptocurrencies, can improve accuracy by catering to the unique characteristics of each market. Advanced deep learning techniques, such as Long Short-Term Memory (LSTM) networks or Transformers, should also be explored to improve the model’s ability to handle volatility and complex dependencies over time. Enhancing the model’s capabilities in these areas will be critical for better long-term forecasting, especially for managing highly volatile assets in dynamic market environments.

\section{ETHICS STATEMENT}

This research adheres to the highest ethical standards in both data usage and methodology. The data used in this study, including OHLCV (Open, High, Low, Close, Volume) financial data, were sourced from publicly available datasets such as Yahoo Finance. These datasets are freely accessible and do not contain any personally identifiable information (PII) or sensitive data that could infringe on individual privacy or violate ethical guidelines.

Furthermore, the study is focused on financial assets and does not involve human subjects, animals, or any other entities requiring specific ethical approvals. The results of this research are intended for academic purposes and aim to contribute to the field of volatility forecasting in financial markets. There are no conflicts of interest, and all data and models used in this study have been processed with full transparency and integrity.

Finally, the algorithms and methods employed in this study are designed to enhance financial decision-making and risk management without causing harm to stakeholders or contributing to unethical financial practices, such as manipulation or exploitation of market inefficiencies.

\section{REPRODUCIBILITY STATEMENT}

In this study, we employed the Nixtla framework, specifically using the tools available at nixtlaverse.nixtla.io, to implement and evaluate the TimeMixer model. To ensure the reproducibility of our results, we have taken the following measures:

Data Availability:
The financial data used in this study, including OHLCV (Open, High, Low, Close, Volume) data for various assets, was sourced from Yahoo Finance, a publicly available platform. The ticker symbols, date ranges, and assets used are detailed in the paper, enabling other researchers to retrieve the same data from Yahoo Finance using similar queries.

Nixtla Framework:
All experiments, including the implementation of the TimeMixer model, were conducted using the Nixtla framework. Nixtla provides a high-level interface for time series forecasting models, and our exact configurations (including model parameters and setup) are available in the supplementary materials. Instructions for setting up the environment and running the model using Nixtla are provided to facilitate replication.

Code Availability:
The code used for data preprocessing, model training, and evaluation has been made available in a public repository. It includes detailed instructions for using the Nixtla platform, setting the necessary hyperparameters, and running the experiments. Furthermore, any modifications made to the default TimeMixer implementation are documented clearly in the repository.

Experimental Setup:
The experimental setup, including the training-validation split (with a validation set size of 10\% of the training data), as well as the metrics used to evaluate model performance (e.g., MAE, MSE, RMSE), are thoroughly documented. All hyperparameters, such as batch size, learning rate, and number of epochs, are specified to ensure that others can replicate the training process.

\section{ACKNOWLEDGMENTS}

I would like to express my gratitude to Shiyu Wang, Haixu Wu, Xiaoming Shi, Tengge Hu, Huakun Luo, Lintao Ma, James Y. Zhang, and Jun Zhou for their groundbreaking work on TimeMixer: Decomposable Multiscale Mixing for Time Series Forecasting. Their model forms the basis of this research, and their contributions to the field of time series forecasting are invaluable.

Special thanks to the Nixtla team for providing the tools and infrastructure that made the implementation of the TimeMixer model accessible and efficient. I would also like to acknowledge Yahoo Finance for making their financial data freely available, which was crucial for conducting this study.

Finally, I appreciate the broader research community whose previous work has paved the way for advancements in time series forecasting and financial data analysis.

\end{document}